\documentclass{iau}

\usepackage{amsmath}
\usepackage{graphicx}
\usepackage{multirow}
\usepackage{natbib}
\usepackage{xspace}
\usepackage{etex}

\def\kms{{\text{km\,s}$^{-1}$}\xspace}

\def\Msun{{\rm M$_{\odot}$}\xspace}
\def\Lbol{{\rm L$_{bol}$}\xspace}

\newcommand{\io}[2]{#1\,{\textsc{#2}}}
\def\ergs{{\rm\,erg\,s$^{-1}$}\xspace}

\begin{document}

\lefttitle{Nadejda Blagorodnova}
\righttitle{IAU Symposium 406 $—$ Future landscape of astrophysical transients}

\jnlPage{1}{7}
\jnlDoiYr{2026}
\doival{10.1017/xxxxx}

\aopheadtitle{Proceedings IAU Symposium IAUS406, 2026}
\editors{Hanin Kuncarayakti, Tuomas Kangas, Melina Bersten}

\title{Exploring stellar evolution with 
Luminous Red Novae and Interacting Gap Transients}

\author{Nadejda Blagorodnova}
\affiliation{Institut de Ciències del Cosmos (ICCUB), Universitat de Barcelona (UB), c. Martí i Franquès, 1, Barcelona, 08028, Spain}
\affiliation{Departament de Física Quàntica i Astrofísica (FQA), Universitat de Barcelona (UB), c. Martí i Franquès, 1, Barcelona, 08028, Spain}
\affiliation{Institut d'Estudis Espacials de Catalunya (IEEC), Edifici RDIT, Campus UPC, Castelldefels, 08860, Spain}

\begin{abstract}
Interacting gap transients encompass several families of astrophysical phenomena where eruptive mass loss and interaction play an important role in shaping the observational properties and energetics of the events. Composed of Luminous Red Novae (LRNe), Intermediate Luminosity Red Transients (ILRTs), Luminous Blue Variable (LBV) giant eruptions, and pre-supernova precursors, interacting gap transients offer crucial insights into unstable mass transfer, extreme stellar mass loss, and binary evolution. In these proceedings, I review the observational signatures, progenitors, physical origin, and classification challenges of these heterogeneous phenomena. Finally, I highlight how ongoing and future wide-field time-domain missions will revolutionize this field by compiling complete volume-limited samples, tracking faint pre-outburst precursors, and establishing statistical links between gap transient families and their progenitor populations.
\end{abstract}


\maketitle

\section{Introduction}
Astrophysical transients serve as luminous probes of key phases in stellar evolution. Because these events mark the culmination of distinct evolutionary phases, they provide vital snapshots into the lives of stars across diverse masses, metallicities, and environments. While terminal transients such as supernovae (SNe) and tidal disruption events (TDEs) mark stellar death, non-terminal transients like massive outbursts and stellar mergers offer complementary insights into how stars live, evolve, and interact with each other while still alive.

\begin{figure}[h]
    \includegraphics[scale=.5]{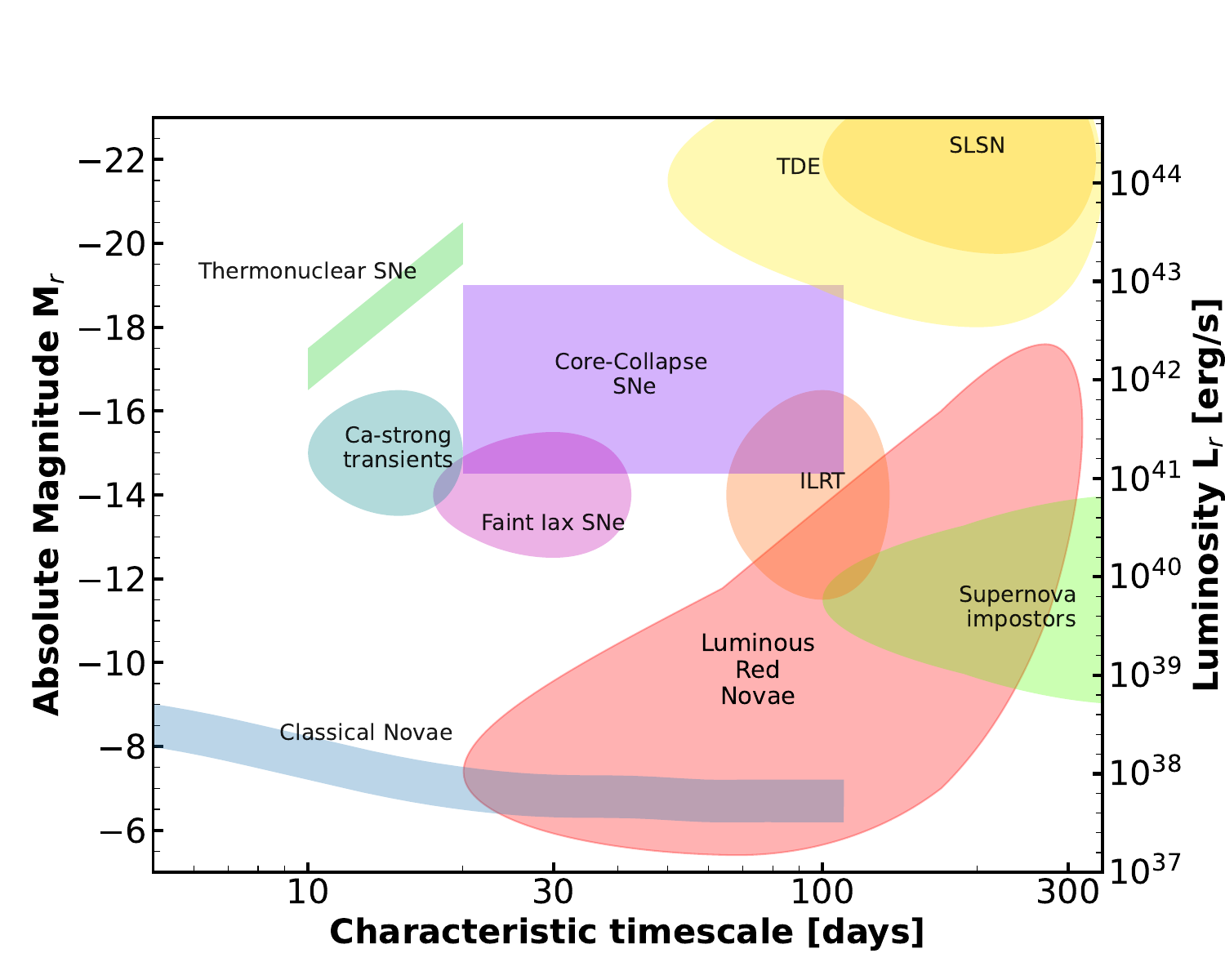}
    \caption{Approximate parameter space of different astrophysical transient families as absolute magnitude (optical luminosity) vs. duration.}
    \label{fig:parameter_space}
  \end{figure}

Historically, the optical transient landscape was divided into two distinct regimes: luminous terminal supernova explosions and fainter non-terminal novae. Over the past two decades, wide-field surveys have systematically filled this apparent luminosity gap with intermediate-luminosity events, broadly termed ``gap transients'' \citep{Kasliwal2012PASA} or Intermediate Luminosity Optical Transients \citep[ILOTs;][]{Soker2016MNRAS} This population encompasses diverse physical mechanisms, including faint thermonuclear explosions (such as Type Iax, 02cx-like SNe; \citealt{Foley2013ApJ}), Ca-strong transients \citep{Perets2010Natur,Kasliwal2012ApJ}, low-luminosity core-collapse SNe \citep{Turatto1998ApJ,Pastorello2009MNRAS}, and interacting, hydrogen-rich transients. The latter group, often designated as ``interacting gap transients'' \citep{Cai2022Universe}, is spectroscopically defined by narrow ($\lsim$1500\,\kms) emission lines of hydrogen and low-ionization species (e.g., \io{He}{i}, \io{Ca}{ii}, \io{Na}{i}, \io{Fe}{ii}, \io{Ti}{ii}, and \io{Ba}{ii}). Figure~\ref{fig:parameter_space} shows the distribution of these three primary classes: Luminous Red Novae (LRNe), Intermediate Luminosity Red Transients (ILRTs), and Supernova Impostors, including giant Luminous Blue Variable (LBV) eruptions and SN precursors. Spanning peak luminosities of $10^{37} - 10^{42}$\ergs, and durations of weeks to months, they bridge the observational gap between novae and SNe and form the core focus of this proceedings review.

\section{Luminous Red Novae - LRNe}

Luminous Red Novae (LRNe) are the high-luminosity extragalactic counterparts of Galactic red novae, historically exemplified by Milky Way transients such as V4332\,Sgr \citep{Martini1999}, V838\,Mon \citep{Munari2002}, V1309\,Sco \citep{Mason2010}, and OGLE-2002-BLG-360 \citep{TylendaBLG}. While the physical nature of these eruptions was initially debated, pre-outburst photometry of V1309\,Sco---which revealed an eclipsing binary undergoing exponential orbital decay before the eruption---provided definitive evidence linking red novae to common envelope evolution in binary systems, stellar mergers, and energetic mass ejection \citep{PaczynskiCE,SokerTylenda2006,Tylenda2011}. To date, approximately 30 LRNe have been reported in the literature, exhibiting remarkable diversity in light-curve morphology and overall energetics, with peak luminosities in the $10^{38}-10^{42}$
\,\ergs range). For comprehensive reviews, see \cite{KaminskiBlagorodnova2026} regarding observational properties and physical interpretation of LRNe and \cite{SchneiderAnnRev} for a broader overview of stellar mergers.

\subsection{LRNe progenitors}

The properties of LRNe progenitor systems can be characterized using archival pre-outburst photometry obtained years before the outburst, when the binary is in ``quiescence''. Multi-wavelength observations of $\sim$20 nearby LRNe have constrained progenitor effective temperatures and luminosities, yielding estimates of stellar mass and evolutionary state \citep[see sample in][]{KaminskiBlagorodnova2026}. Remarkably, most identified progenitors are quickly expanding post-main-sequence yellow giants or supergiants traversing the Hertzsprung gap \citep{Blagorodnova2017}.

Within the stellar merger framework, LRNe mark the culmination of unstable mass transfer in binary systems. Although several mechanisms can draw angular momentum from an orbit, this unstable regime is most frequently triggered when the donor star undergoes thermal expansion and fills its Roche lobe. In a typical scenario, the primary star exhausts core hydrogen and leaves the main sequence. As it rapidly expands and cools at nearly constant luminosity, it traverses the Hertzsprung gap toward the red giant branch. Because massive stars predominantly reside in tight binary systems with small orbital separations \citep{MoeDiStefano2017}, the donor star's thermal expansion rapidly leads to binary interaction. The donor fills its Roche lobe and initiates runaway mass transfer, rapid orbital decay through energy and angular momentum loss, and ultimately a stellar merger. LRNe thus provide a unique observational diagnostic for the end of unstable mass transfer, which leads to common envelope evolution and, if the merger is ultimately avoided, the formation of compact binary systems \citep{PaczynskiCE,Ivanova2013Rev}. Consequently, characterizing LRN progenitors and their outbursts is essential for mapping dynamical interactions in close binaries.

For a handful of LRNe in the Milky Way and nearby galaxies, multi-year pre-explosion data provide valuable insight into the final evolutionary stages before coalescence. In this sample, progenitor systems display a slow brightening of 2--4 magnitudes over the 2--10 years preceding the main nova event \citep[e.g.,][]{Blagorodnova2017,Blagorodnova2020,Pastorello2021_2019zhd,Reguitti2026}. This precursor emission, which was also observed for the eclipsing binary progenitor of V1309\,Sco, has been associated with the continuous mass loss through the outer Lagrange L2/L3 points in the binary system \citep{Pejcha2017}. As mass loss rapidly escalates, the expanding outflows shape a self-interacting spiral pattern, producing shock-heated gas that substantially increases the system's luminosity. Because L2/L3 outflows efficiently remove orbital angular momentum, the orbit rapidly decays, driving the binary toward coalescence \citep{MacLeod2020ApJ}. Recent studies indicate that as the companion plunges into the donor’s outer envelope, the thermalization of drag forces can also provide an extra energy source, further elevating the system's luminosity in the final pre-merger years \citep[][Karambelkar in prep.]{Bronner2024}.

A notable exception to yellow supergiant progenitors is the recent LRN AT2025abao in M31, marking the fourth LRN discovered in this galaxy since 1989 and the first with a confirmed AGB progenitor \citep[7$\pm$2\Msun;][]{Karambelkar2025ApJ,Reguitti2026}. Prior to the main event, its precursor emission was observed for $\sim15$\,years as a slowly evolving optical and infrared transient, leading to an initial interpretation as a possible full common envelope ejection event (forming a tight binary pair). Months after its initial discovery, however, AT2025abao erupted as an optical plateau-type LRN. This behavior directly challenges the expectation that merger outbursts involving highly extended, dusty donors should yield prolonged transients like OGLE-2002-BLG-360 \citep[having a low-mass red supergiant progenitor;][]{TylendaBLG} or dust-enshrouded IR-only transients \citep{Jencson2019SPIRTES}.

\subsection{Observational signatures}
\begin{figure}[h]
  \begin{center}
    \includegraphics[scale=.6]{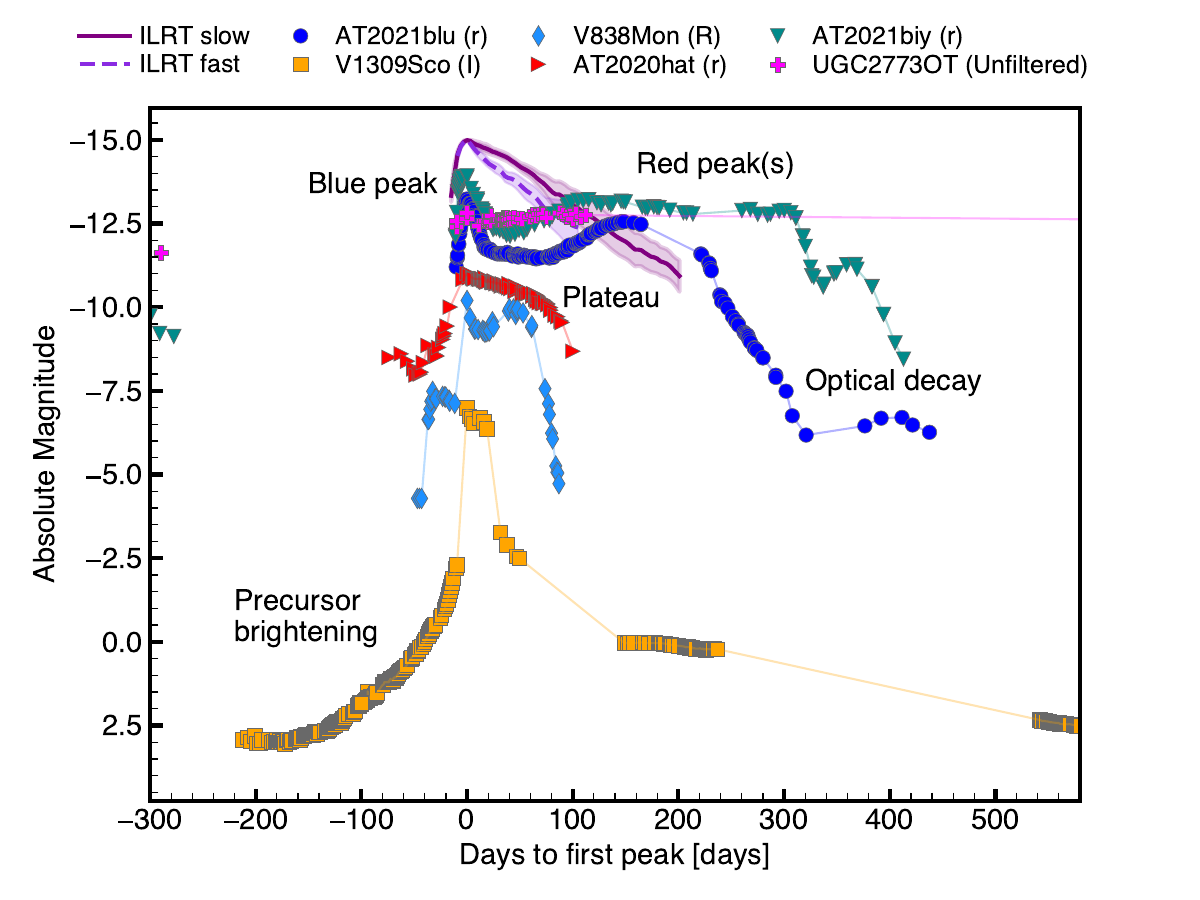}
    \caption{Markers represent different LRNe, according to the legend. The warm colours red and orange represent LRNe of the ``plateau'' group V1309Sco \citep{Mason2010} and AT2020hat \citep{Pastorello2021AA}. The cold colours represent LRNe of the ``risers group'' like V838Mon, AT2021blu \citep{Pastorello2023}, and AT2021biy \citep{Cai2022}. Magenta shows the slowly evolving lightcurve of the decade-long LBV UGC2273-OT \citep{Smith2016MNRAS}. The photometric band shown in the plot for each lightcurve is indicated in parentheses in the legend. The solid and dashed lines represent the templates from \cite{Valerin2025AA_ILRTphot} for fast and slow ILRT, scaled to $-15$ peak magnitude, with the shaded area representing the uncertainty of the template. }
    \label{fig:ligthcurves}
      \end{center}
  \end{figure}

Beneath their apparent diversity, the light curves of LRNe appear to fall into two main morphologies: a ``plateau group'' characterized by extended optical plateaus, and a ``risers group'' defined by one or more prominent re-brightening episodes \citep{KaminskiBlagorodnova2026}, as shown in Fig.~\ref{fig:ligthcurves}. Light curves across both morphological groups typically exhibit an initial hot, blue peak with early photospheric temperatures reaching up to 18,000\,K \citep{Cai2022}, which represents the initial release of thermal energy of the fast material ejected during the merger. Following the initial peak, the light curves in the ``plateau group'' settle into an almost constant luminosity in the optical $r$-band with durations between a month (like V1309\,Sco) and up to 100\,days. This group spans a wide range of peak absolute magnitudes, from $M_V=-7\pm1.4$ for V1309\,Sco up to $M_R=-14.3$ for AT2007sv \citep{ReguittiIR}. On the contrary, the ``risers group'' generally encompasses objects brighter than $-12$\,mag (with the exception of the Galactic transients V838\,Mon and OGLE-2002-BLG-360) and displays longer durations of $100-300$\,days. For systems with well-characterized progenitors, outburst peak brightness and duration strongly correlate with progenitor mass, as more massive systems produce brighter, longer-lived eruptions \citep[e.g.,][]{Kochanek2014,Pastorello2019review,Blagorodnova2021}. 

Over the past decade, significant theoretical effort has focused on modeling LRN light curves. While pure hydrogen-recombination models successfully reproduce the plateau-like morphology of lower-luminosity events \citep{IvanovaScience,MatsumotoMetzger}, applying them to the most luminous ``risers'' requires unphysically large ejecta masses that exceed total progenitor envelope constraints. This discrepancy indicates that interaction with pre-existing circumstellar material is necessary to thermalize the kinetic energy of the ejecta, driving the extended secondary red peaks \citep{MetzgerPejcha2017,Kirilov2025ApJ}. In all cases, more massive donor stars are expected to yield larger ejecta masses at higher velocities, enhancing both recombination power and shock heating.

True to their ``red'' designation, LRNe exhibit exceptionally low effective temperatures towards the end of the optical lightcurve decay, well below 3000\,K, resembling late M-type stars. This cooling environment enables rapid molecule and dust formation in the expanding ejecta, which subsequently re-radiates optical emission into IR wavelengths. Consequently, while the optical emission fades within a few months, long-lived IR emission remains detectable for several years post-outburst \citep{Blagorodnova2020,ReguittiIR}.

\begin{figure}[h]
  \begin{center}
    \includegraphics[width=\textwidth]{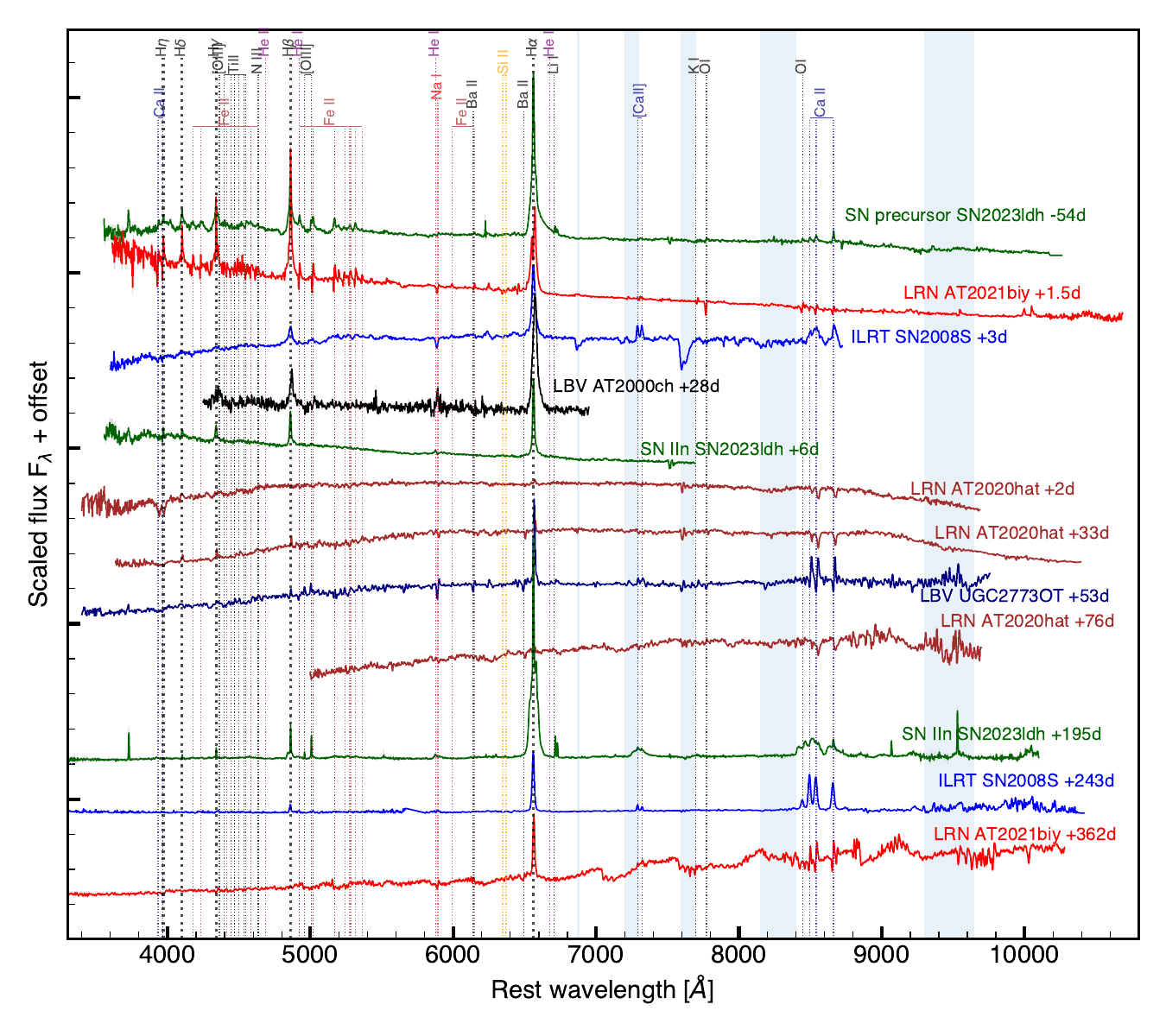}
    \caption{Spectroscopic sequence of interacting Gap Transients and the precursor supernova type IIn SN2023ldh. At early times, the transients share similar characteristics, showing strong hydrogen emission lines, \io{Fe}{ii}, \io{Ca}{ii}, and \io{He}{i} among others. The data are taken from the following sources SN2023ldh (green) \citep{Pastorello2025AA}, AT2021biy (red) \citep{Cai2022,Karambelkar2023ApJ}, SN2008S (blue) \citep{Botticella2009MNRAS}, LBV AT2000ch (black) \citep{Wagner2004PASP}, LRN AT2020hat (brown) \citep{Pastorello2021AA}, LBV UGC2773OT TNG/DOLORES (navy) (A. Pastorello 2016, private communication). Epochs are given relative to maximum or to the discovery date.}
    \label{fig:spectralsequence}
      \end{center}
  \end{figure}

In line with their photometric evolution, the spectroscopic evolution of LRNe tracks a steady transition toward lower ionization states and cooler effective temperatures (see Fig.~\ref{fig:spectralsequence}). During the initial hot peak, the transients usually display strong hydrogen emission lines, often broadened by electron scattering \citep{Sneppen2026arXiv}. High-resolution observations at early phases often reveal P-Cygni profiles in \io{Fe}{ii} and \io{Ti}{ii}, alongside \io{Na}{i}, \io{Ba}{ii}, \io{Si}{ii}, and \io{Ca}{ii}. Past the peak, hydrogen emission fades rapidly while a dense absorption ``forest'' of neutral and low-ionization species emerges, causing the spectrum to temporarily resemble an F-type star. As the expanding ejecta continue to cool toward the end of the optical plateau, molecular bands form and progressively become stronger. Later-time infrared spectroscopy confirms that the remnant pseudo-photosphere resembles an inflated M-type giant star, dominated by deep molecular absorption features of water vapour, TiO, VO, CO, and in some cases, AlO compounds \citep[e.g.,][Gomez-Muñoz in prep.]{Banerjee2005,Banerjee2003ApJ,Blagorodnova2021}.

\subsection{Remnants}

While rapid dust formation causes extragalactic LRNe to fade quickly at optical wavelengths, the central remnants of Galactic red novae remain visible in the optical \citep[except OGLE-2002-BLG-360; see][]{Steinmetz2025}, albeit heavily contaminated by circumstellar emission and absorption. Resolved millimeter imaging reveals that ejected merger debris continuously interacts with pre-existing circumbinary and circumstellar gas \citep{Steinmetz2024}, shaping the outflow into a bipolar geometry \citep{Kaminski2018AA} analogous to planetary nebulae \citep{Jones2017NatAs}. For instance, the central star of V838\,Mon physically resembles an inflated M-type supergiant; 24 years after its merger, it has recently initiated a pulsation cycle akin to Mira-type variables \citep{Kaminski2026arXiv}. In contrast, the central star of V4332\,Sgr is completely obscured by an edge-on dusty torus, though its underlying photospheric spectrum can be successfully recovered via polarized light scattered off surrounding dust grains \citep{Kaminski2013,KaminskiBlagorodnova2026}.

For extragalactic LRNe, substantial dust production completely obscures the central star at optical wavelengths, leaving it observable only via reprocessed mid-infrared (MIR) emission. When incorporating the flux contribution from longer wavelengths, the bolometric luminosity of the remnant at $\sim$3\,years post-outburst is only one to two orders of magnitude fainter than during peak (Gomez-Muñoz in prep.), indicating that the merger-disturbed star is slowly relaxing back toward hydrostatic equilibrium. As the ejected gas cools, rapid dust condensation occurs: James Webb Space Telescope (JWST) observations have shown that $10^{-3} - 10^{-5}$\,\Msun of oxygen-rich dust forms in LRN remnants within the first decade following the outburst \citep{ReguittiIR,KarambelkarJWST,Steinmetz2025}. Future infrared surveys will map these dust masses and compositions across different progenitor channels, providing critical empirical inputs to advance dust formation simulations \citep[e.g.,][]{BermudezBustamente2024,Mu2026arXiv} and quantify the overall contribution of LRNe to the interstellar dust budget \citep{KarambelkarJWST}.

\section{Intermediate Luminosity Red Transients - ILRTs}

Intermediate Luminosity Red Transients (ILRTs) represent a distinct class of gap transients, represented by the iconic astrophysical events NGC300-OT and SN\,2008S \citep[][and references therein]{Berger2009ApJ,Botticella2009MNRAS}. Although historically designated in the literature as Intermediate Luminosity Optical Transients (ILOTs), the community has broadly adopted the term ILRT for this specific class, reserving ``ILOT'' as an umbrella term for the broader gap transient population \citep{Soker2016MNRAS}.

\subsection{Progenitors}

A key defining feature of ILRT progenitors is their severe dust obscuration, characterized by low effective dust re-processing temperatures (T$<$500\,K) and high bolometric luminosities of $(4-5)\times 10^4$\,\Lbol, matching the expectations for 8$–$12\,\Msun super-AGB and EAGB stars \citep{Prieto2008}. Because such extreme circumstellar extinction is exceedingly rare across stellar populations in similar host galaxies, \citet{Thompson2009ApJ} inferred that this dust-enshrouded phase represents a short-lived evolutionary state enduring for $\lsim $10,000\,years. An additional demographic constraint was provided by M85OT2006-1, an ILRT originally classified as an LRN \citep{Kulkarni2007Natur}. Non-detections of young stars in the transient's vicinity enforce an age limit of $\gsim$50\,Myr, favoring a progenitor mass $<$7\,\Msun
  \citep{Ofek2008ApJ}. Furthermore, constraints from the host galaxy's mean stellar population age ($\sim$1.6\,Gyr) point to an even lower progenitor mass of $\lsim$2\,\Msun.

Although the progenitor nature of ILRTs remains debated, recent JWST observations of the ILRT AT2019abn at +1421days post-outburst confirm the presence of Polycyclic Aromatic Hydrocarbons (PAHs) \citep{Rose2025ApJ}, echoing earlier detections in NGC300-OT \citep{Prieto2009ApJ}. Reminiscent of environments around low-mass ($<$4\,\Msun) post-AGB and Herbig Ae/Be stars, these molecular features reveal the presence of carbon-rich dust in the circumstellar medium, favouring lower-mass progenitors: in standard evolutionary models of massive AGB stars, Hot-Bottom Burning (HBB) prevents carbon enrichment by efficiently processing dredged-up carbon into nitrogen through CNO cycle at the base of the convective envelope. Hence, the consistent detection of carbonaceous dust challenges the super AGB progenitor scenario, unless C-rich photospheres can briefly form in massive AGB stars under specific conditions \citep{Marigo2007}.
 
\subsection{Observational signatures}

Compared to LRNe, the photometric evolution of ILRTs is remarkably uniform, displaying peak visual absolute magnitudes in the range of $-12 \lsim M_V \lsim -15$ and optical luminosities of $10^{40} - 10^{41}$\,\ergs. Their light curves are characterized by a steep initial rise followed by a slow, linear decline lasting 100$-$200\,days, before transitioning to a steeper drop and a final, shallower tail. Throughout their evolution, ILRTs exhibit moderate color temperatures (5000$-$8000\,K) and minimal color evolution ($B-V$ rarely evolves redder than 1\,mag at late times), which uniquely distinguishes them from the progressively reddening light curves of LRNe. A recent study by \cite{Valerin2025AA_ILRTphot} has identified distinct slow- and fast-decaying subpopulations within the ILRT class, as shown in Fig.~\ref{fig:ligthcurves}. Across both subtypes, a distinct cool IR emission component develops in the SED at $\sim$100\,days post-peak, complementing the slow decline of the primary optical continuum.

Spectroscopically, ILRTs are defined by prominent hydrogen emission lines featuring broadened bases shaped by electron scattering (e.g., SN\,2008S; Fig.~\ref{fig:spectralsequence}). The optical spectra consistently exhibit strong \io{Ca}{ii} transitions, marked by deep \io{Ca}{ii} H\&K absorption and intense \io{Ca}{ii} near-infrared triplet emission. A key diagnostic distinguishing ILRTs from other gap transients is the ubiquitous presence of strong forbidden [\io{Ca}{ii}] at $\lambda\lambda$7291, 7323. High-resolution spectra of NGC300-OT also allowed to the identification of \io{Na}{i} absorption, \io{He}{i}, and the presence of permitted and forbidden \io{Fe}{ii}, and \io{O}{i} \citep{Berger2009ApJ,Valerin2025AA_ILRT_spec,Mason2026}. In contrast to the deep molecular absorption bands that dominate late-time LRNe spectra, ILRTs never develop detectable molecular features in either optical or infrared regimes, maintaining spectra dominated by persistent hydrogen and \io{Ca}{ii} emission into late phases.

\subsection{Physical nature}

The physical nature of ILRTs remains debated. Long-term multi-year infrared follow-up reveals that their total luminosities eventually fade well below their pre-outburst progenitor levels \citep{Adams2016MNRAS,Rose2025ApJ,Valerin2025AA_ILRTphot}. Given that dust opacity becomes inefficient at mid-IR wavelengths, this permanent disappearance strongly points toward a terminal explosion. A leading hypothesis attributes ILRTs to electron-capture supernovae (ECSNe) in super-AGB progenitors \citep[e.g.,][]{Botticella2009MNRAS,Pumo2009ApJ}. Rather than undergoing core oxygen ignition to form an iron core, the degenerate O–Ne–Mg core of super-AGB stars is expected to undergo rapid electron capture on $^{24}$Mg and $^{20}$Ne atoms. The resulting loss of electron degeneracy support reduces the core Chandrasekhar limit, initiating collapse into a low-energy explosion \citep[$\simeq 10^{50}$\,erg, see][]{Miyaji1980PASJ,Kitaura2006,Podsiadlowski2004}. Owing to low kinetic energies and extremely low \io{Ni}{} production ($\simeq 10^{-3}$\,\Msun), these events can reproduce total radiated energies of ILRTs of $10^{47} - 10^{48}$\,erg. Interestingly, the estimated volumetric rate of ILRTs of $2.6^{+1.8}_{-1.4} \times 10^{-6} \textrm{Mpc}^{-3} \textrm{yr}^{-1}$ \citep{Karambelkar2023ApJ} would be an order of magnitude higher than the upper limit for ECSN identified from nebular spectroscopy low-luminosity core-collapse SNe, with a rate of $\leq (5-8) \times 10^{-7} \textrm{Gpc}^{-3} \textrm{yr}^{-1}$ \citep{Das2026PASP}. If ILRTs could account for ``missing'' ECSN population, this would help to alleviate the currently narrow mass window ($\Delta \textrm{M}_{\textrm{sAGB}} \lsim  0.02-0.06$\,\Msun) predicted for super-AGB stars to undergo electron capture-driven collapse. Nevertheless, attributing ILRTs to ECSNe poses a major spectroscopic challenge: a distinct physical mechanism is needed to reconcile their unusual nebular signatures—dominated by persistent hydrogen and forbidden calcium lines—with the oxygen- and iron-rich nebular emission typical of canonical core-collapse events.

\section{Supernova Impostors and Luminous Blue Variables}

Non-terminal outbursts that mimic CSM-interacting supernovae within the luminosity gap have long been broadly classified as ``supernova impostors'' \citep[e.g.,][]{Foley2011ApJ,VanDyk2012,Kochanek2012ApJ}. As the classification scheme evolved to isolate LRNe and ILRTs into distinct physical families, the remaining impostor population primarily comprises two distinct phenomena: pre-explosion eruptive ``precursors'' preceding SN IIn/IIP explosions, as for SN2009ip, SN2020tlf, SN20023ldh \citep[e.g.,][]{Smith2010AJ,Margutti2014ApJ,JacobsonGalan2022ApJ,Pastorello2025AA}, and giant eruptive mass-loss events from Luminous Blue Variable (LBV) stars, analogous to $\eta$ Car and P Cygni. Readers are directed to comprehensive reviews detailing the physical drivers and observational properties of both classes \citep{Smith2014ARAA,Smith2011_LBVs,Langer2012ARA,Smith2026enap}.

\subsection{Observational signatures}

LBVs' photometric evolution is characterized by multi-year evolution timescales \citep{Smith2016MNRAS} and recurring eruptive episodes \citep{Aghakhanloo2023a}. Supernova precursors can also exhibit diverse temporal behavior, ranging from one-month to steady, long-duration brightenings of the progenitor star. These pre-explosion outbursts are especially frequent in the final stages of Type IIn supernova progenitors, where approximately 25\% display months-long precursors brighter than $-13$\,M${_r}~$ during the final three months leading up to core collapse \citep{Strotjohann2021ApJ}.

At early phases, the optical spectra of supernova impostors exhibit narrow hydrogen emission lines ($\lsim1500$\,\kms), closely resembling CSM-interacting Type IIn supernovae (see Fig.~\ref{fig:spectralsequence}). In some cases, strong emission for \io{Fe}{ii} and \io{Ti}{ii} is also detected, analogous to the most energetic LRNe. This substantial overlap in peak luminosity, duration, and early spectral morphology poses a significant observational challenge for prompt transient classification. As demonstrated in Fig.~\ref{fig:spectralsequence}, distinguishing between LRNe, ILRTs, LBVs, and pre-explosion SN precursors is often unfeasible from standard low-resolution, low-signal-to-noise classification spectra. Adding to this ambiguity, the expanding wind envelopes of LBV eruptions can cool over several months to form opaque, F-type pseudo-photospheres, temporarily creating strong spectral degeneracies with post-peak LRNe.

\subsection{Physical nature}

The physical mechanisms governing LBV eruptions remain a subject of ongoing research. Originally conceptualized as super-Eddington winds arising from radiation-dominated envelope instabilities in high-mass stars ($>$30\,\Msun), the observed progenitor mass range has now expanded to include lower masses \citep[e.g. UGC2773-OT;][]{Smith2010AJ}. Among LBVs, different mechanisms are likely to be at play. Whereas standard S\,Doradus cycles and giant eruptions cool to F-type pseudo-photospheres at maximum light (with minimum variation in their bolometric luminosities), the so-called ``hot LBV eruptions'', having early Wolf-Rayet (Ofpe/WN9) progenitors, keep their high temperatures through the outburst. A plausible physical scenario would be related to H-poor (but not H-free) LBV stars as they approach the Wolf-Rayet phase, offering a connection to H-poor precursor eruptions observed before Type\,Ibn supernovae \citep[e.g.,][]{Pastorello2015MNRAS,Smith2012MNRAS}.

A further major development in understanding LBV activity is the recognition of strictly periodic and quasi-periodic eruptive behavior, emphasizing the role of binary interactions over purely intrinsic stellar instabilities. A clear representative is $\eta$ Car, where a 5.52-year periastron cycle in a highly eccentric orbit ($e\simeq$0.9) regularly experiences wind-wind collision \citep{Okazaki2008}. Other examples are SN2000ch and AT2016blu, exhibiting periodic eruptions every $\sim$200\,days and $\sim$113 days, respectively \citep[e.g.,][]{Pastorello2010,Aghakhanloo2023a,Aghakhanloo2023b}. In these systems, repeated eruptive flares are driven by violent periastron encounters, demonstrating that binary dynamics play a critical role in triggering giant mass-loss events in massive stars. In addition, the apparent isolation of LBVs as compared to WR stars can also support their origin as mass gainers or stellar merger products, thus retaining their H-rich envelopes until their death as SN type IIn \citep{Justham2014,Deman2024}.

Eruptive SN precursors have been explained through several physical scenarios. The primary channel involves the propagation of gravity waves from the stellar core into the envelope, as vigorous core convection during the final nuclear burning can arise due to super-Eddington fusion luminosity. These are expected to occur in years to days before collapse for O/Ne and Si burning, respectively \citep[e.g.,][]{Quataert2012,Fuller2017}. Alternative mechanism proposals invoke hydrodynamic instabilities arising during late-stage nuclear shell burning, which inject explosive energy waves into the outer envelope \citep{Smith2014ApJ,Yoshida2021}. Additionally, binary interactions and mass-transfer episodes---such as orbital decay or eruptive envelope stripping---can drive substantial mass ejection prior to explosion \citep[e.g.,][]{Chevalier2012,Sun2020}. This produces a dense circumstellar environment around the progenitor, which later will define the geometrical and radiative evolution of the expanding supernova ejecta.

\section{Open questions and future research}

Despite a growing observational sample, several fundamental questions surrounding interacting gap transients remain unresolved. In the case of LRNe, the connection between mass loss, orbital decay, precursor emission, and spectroscopic evolution is not clear. To date, V1309\,Sco remains the only system where exponential orbital decay was explicitly detected alongside pre-merger precursor brightening; for most extragalactic events, the precursor photometric observations lack the cadence and depth required to impose stringent constraints. It therefore remains undefined how mass and angular momentum are lost from the binary system prior to dynamic plunge-in, and how the resulting circumbinary medium shapes the optical light curve. It is still under question whether the progenitor primary mass, evolutionary stage, and binary mass ratio uniquely determine LRN light-curve properties, or conversely, to what extent progenitor parameters can be uniquely estimated from observed light curves alone. While multi-dimensional simulations continue to advance our theoretical framework \citep{MetzgerPejcha2017,MacLeodLoeb-preCE,MatsumotoMetzger,Kirilov2025ApJ}, quantitative light-curve fits have been only attempted for a limite number of benchmark systems \citep[e.g.,][]{NandezV1309,Pejcha2017,HatfullIvanova2021,ChenIvanova2024}.

Looking toward the broader transient landscape, a compelling challenge lies in connecting single and binary evolution pathways to observations of progenitors and the outbursts LRNe, ILRTs, LBV giant eruptions, pre-explosion precursor flares, and core-collapse supernovae \citep[e.g.][]{Blagorodnova2021,Wavasseur2026arXiv,Prieto2008, Laplace2021,Laplace2026}. The high frequency of binarity (and higher multiplicity) among massive stars strongly supports a paradigm where binary-driven non-terminal outbursts frequently dictate the late-stage evolution and eventual fate of massive stars \citep{Ercolino2024}. A prime example of this is the nearby SN\,1987A, whose blue supergiant progenitor is widely accepted to have been a $\sim10^4$-year-old merger remnant \citep{Podsiadlowski1992ApJ}.

The ongoing Euclid survey, the initiation of the Vera C. Rubin Observatory's Legacy Survey of Space and Time (LSST), and the upcoming Nancy Grace Roman Space Telescope (among other upcoming missions) signal a bright future across the electromagnetic spectrum. The high spatial resolution, deep multi-band imaging, and regular cadences of these missions will allow us to identify and monitor lower-luminosity pre-outburst events and their progenitor stars well before dynamic eruption, offering the unprecedented capability to forecast upcoming stellar mergers and supernova explosions. Preliminary searches conducted on shallower time-domain surveys highlight the feasibility of this approach \citep{Tranin2025}. Concurrently, the dramatically expanded survey volumes will rapidly build large, uniform samples, providing statistically robust constraints on transient rates, light-curve properties, and circumstellar environments. Linking these observational signatures back to initial single- and binary-star parameters will enable us to map and quantify multiple stellar evolutionary channels for the massive star population.

\section{Acknowledgements}
I would like to thank Andrea Pastorello for making available the observational data compiled by their group. I acknowledge the use of the IA tool Gemini to help improve the clarity and flow of my own text. I also used it to reformat LaTeX photometry tables published in each paper into other machine-readable formats to create the plots for this review. I acknowledge being funded by the European Union (ERC, CET-3PO, 101042610). Views and opinions expressed are, however, those of the author(s) only and do not necessarily reflect those of the European Union or the European Research Council Executive Agency. Neither the European Union nor the granting authority can be held responsible for them.
I acknowledge financial support from grant CEX2024-001451-M funded by MICIU/AEI/10.13039/501100011033, and the research project PID2024-155585NA-I00 funded by MICIU/AEI /10.13039/501100011033.

\bibliographystyle{iaulike}
\bibliography{references}

\end{document}